\documentclass[10pt,a4paper,twocolumn,superscriptaddress]{revtex4}

\usepackage{textcomp}
\usepackage{color}
\usepackage{amsmath}
\usepackage{mathrsfs}
\usepackage{graphicx}
\usepackage{subfig}
\usepackage{epsfig}
\usepackage{physics}
\usepackage{hyperref}

\usepackage[squaren,Gray]{SIunits}

\begin{document}

\title{High peak-to-background-ratio solitons in a coherently-driven active fiber cavity}

\author{Nicolas Englebert}
\email{nicolas.englebert@ulb.be}
\author{Carlos Mas Arabí}
\author{Simon-Pierre Gorza}
\author{Fran\c{c}ois Leo}

\affiliation{Service OPERA-\textit{Photonique}, Universit\'e libre de Bruxelles (U.L.B.), 50~Avenue F. D. Roosevelt, CP 194/5, B-1050 Brussels, Belgium}

\begin{abstract}
 We demonstrate that the peak-to-background ratio of driven solitons can be greatly improved by harnessing the cavity detuning. We use a driven fiber laser pumped below the lasing threshold to increase the finesse and excite solitons in a very wide region of detunings. When driving a 50~m long fiber cavity close to the anti-resonance condition, we excite sub-800~fs solitons with a peak-to-background ratio over 30000. The experimental results are in good agreement with simple theoretical models describing the soliton peak power and the background power.
 
\end{abstract}

\maketitle
Temporal cavity solitons (CSs) are short optical pulses that propagate indefinitely in driven Kerr resonators. They have been observed in fiber cavities~\cite{leo_temporal_2010,jang_ultraweak_2013,englebert_temporal_2021}, integrated resonators~\cite{herr_temporal_2014,yi_soliton_2015} and free-space cavities~\cite{lilienfein_temporal_2019}. 
CSs constitute a subset of the wider class of dissipative solitons\,\cite{akhmediev_dissipative_2005}. Specifically, 
they are coherently driven first-order solitons of the nonlinear Schrodinger equation whose duration and peak power are set by the cavity and driving parameters. In particular, the soliton peak power increases when the resonator is driven out of resonance because the nonlinear phase of the soliton compensates for the linear detuning. The main advantage of driven solitons over other types of dissipative solitons (e.g., laser solitons~\cite{grelu_dissipative_2012}) is that they are phase-locked to the continuous wave driving. This can be harnessed for applications, such as dual-comb spectroscopy~\cite{suh_microresonator_2016} or ranging~\cite{trocha_ultrafast_2018}, where the coherence of the pulse train is important.
However, CSs sit atop a continuous wave background which can be detrimental for applications as it saturates amplifiers or detectors. The background of high repetition rate soliton trains (as in microresonators) can be easily filtered out, but it is more difficult in long resonators\,\cite{englebert_bloch_2023}.
There is hence a trade-off between seamless phase locking and pulse train quality, which we here characterize through the peak-to-background-ratio.
Several techniques to increase the peak-to-background ratio of CSs in fiber resonators have been implemented, including two cavity configurations~\cite{xue_super-efficient_2019} or, more recently, our demonstration of active cavity solitons~\cite{englebert_temporal_2021,MasArabi:22} but the energy stored in the background is still orders of magnitude larger than that of the soliton in these experiments.
Synchronous driving~\cite{anderson_observations_2016,obrzud_temporal_2017,lilienfein_temporal_2019} can be used to increase the \emph{energy} ratio, but pulsed driving adds noise and requires more complex experimental designs~\cite{li_experimental_2020}.
In this letter, by harnessing the cavity detuning in an active fiber cavity, we demonstrate the formation of short, high-peak-power solitons on a very low power continuous-wave background. When driving the cavity close to antiresonance, we obtain a peak-to-background ratio of $3\times 10^4$ which is an 80-fold increase compared to the state of the art~\cite{englebert_temporal_2021}.

We start with a theoretical analysis of the peak-to-background ratio of driven solitons. We consider a cavity of length $L$, with average group velocity dispersion $\beta_2$ and average nonlinear parameter $\gamma$, driven by the field $E_\mathrm{in}=\sqrt{P_\mathrm{in}}$ through a coupler with a transmission coefficient $\theta$. The total cavity loss is $\Lambda = -\ln{T}$, where $T$ is the cavity transmission in the absence of driving ($|E_{n+1}|^2=T|E_{n}|^2$),
corresponding to a finesse $\mathcal{F}=2\pi/\Lambda$. The driving field is detuned from the closest resonance by $\delta = 2k\pi-\beta L$ where $\beta$ is the wavevector at the frequency of the driving laser.
In the regime of high finesse ($\mathcal{F}\gg1)$ and high detuning ($\delta\gg\Lambda$), the envelope of CSs is well approximated by the expression~\cite{wabnitz_suppression_1993,coen_universal_2013}
\begin{equation}\label{eqsol}
E_{s}\approx\sqrt{\frac{2\delta}{\gamma L}}\mathrm{sech}\left(\sqrt{\frac{2\delta}{|\beta_2|L}}t\right)\mathrm{e}^{i\phi},
\end{equation}
where the soliton phase satisfies the condition 
\begin{equation}
\cos{\phi}=\frac{\Lambda}{\pi E_\mathrm{in}}\sqrt{\frac{2\delta}{\gamma L\theta}}.
\end{equation}
CSs coexist with a continuous wave intracavity field. Here we use the high finesse ($\mathcal{F}\gg1)$ and 'out of resonance' ($\delta\neq 2k\pi$) approximation of the Airy distribution instead of the oft-used Lorentzian approximation~\cite{wabnitz_suppression_1993}. The background power in that regime is 
\begin{equation}
P_\mathrm{b}\approx\frac{\theta P_\mathrm{in}}{2(1-\cos{\delta})}.
\end{equation}
Considering that the output pulse train is measured in a drop port,
the peak-to-background power ratio is the same as inside the cavity, i.e
\begin{equation}\label{eqratio}
R = \frac{4\delta(1-\cos{\delta})}{\theta P_\mathrm{in}\gamma L}.
\end{equation}
Solitons exist up to the maximum detuning $\delta^\mathrm{max}=\frac{\pi^2P_\mathrm{in}\gamma L\theta}{2\Lambda^2}$~\cite{wabnitz_suppression_1993}. At that detuning, the soliton is in phase with the driving field ($\cos{\phi}=1$), and its peak power is $P^\mathrm{max}_{s}=\frac{\pi^2P_\mathrm{in}\theta}{\Lambda^2}$. The corresponding ratio is 
\begin{equation}
R(\delta_\mathrm{max}) = \frac{\mathcal{F}^2(1-\cos{\delta_\mathrm{max}})}{2}.
\end{equation}
This simple expression clearly highlights the importance of the cavity finesse for high-quality pulse train generation using driven solitons.
When $\delta^\mathrm{max}$ is an odd multiple of $\pi$, the ratio is maximized to $R^{\mathrm{max}} = \mathcal{F}^2$.

\begin{figure}
    \centering
    \vspace{2mm}
    \includegraphics[scale=1.2]{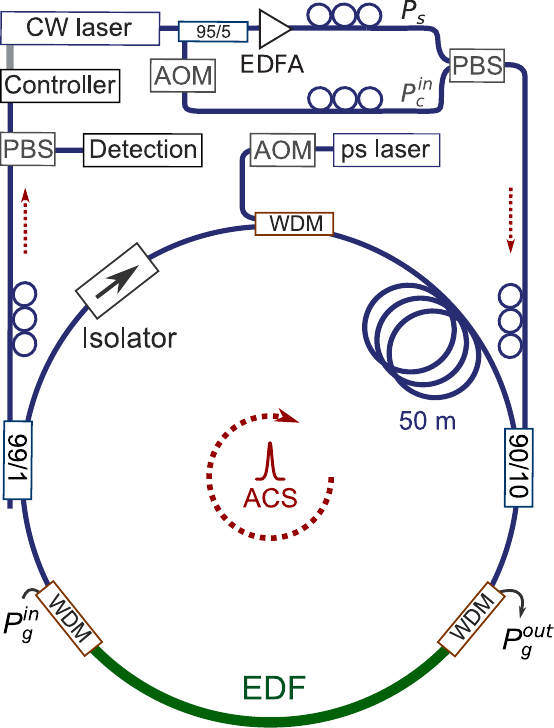}
    \caption{Experimental setup for soliton generation. The solitons are driven through a 90/10 port by a highly coherent laser at 1550~nm. The light is split in two with 5~\% used as a frequency-shifted control signal for interferometric stabilisation. The cavity is ~50~m long and contains an isolator, a 99/1 drop port, a 1550/1535 wavelength multiplexer (WDM), two 1480/1550 WDMs and a 30?~cm long erbium-doped fiber. The solitons are excited using a 1535~m pulse from a mode-locked laser, gated by an acousto-optic modulator (AOM). The gain fiber is pumped at 1480~nm.
    PBS : polarization beam splitter. PC : polarization controler.}
    \label{fig:setup}
\end{figure}
\begin{figure*}
    \centering
    \includegraphics[scale=0.9]{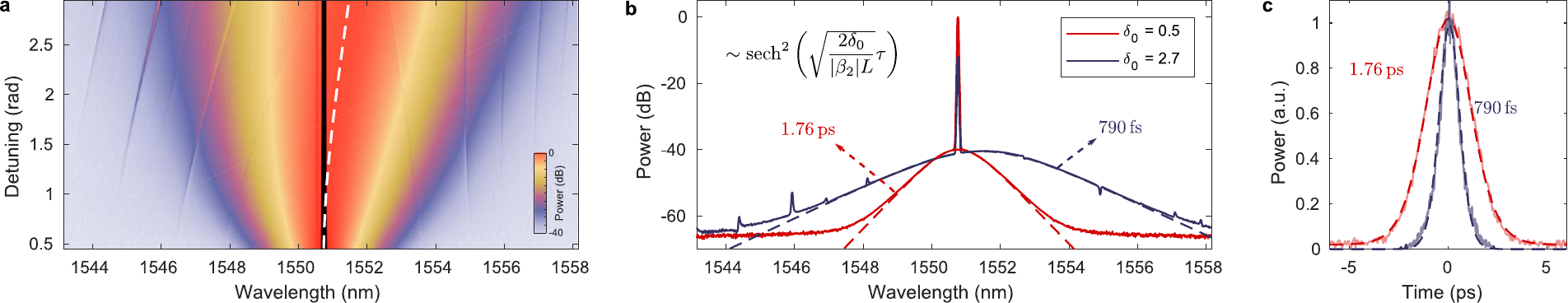}
    \caption{\textbf{a} Experimental optical spectra of the solitons as a function of the detuning. The solid black line indicates the driving wavelength. The dashed white line is the soliton center wavelength. \textbf{b}, Two experimental spectra corresponding to different cavity detunings. The dashed lines correspond to the theoretical scaling law (as shown). \textbf{c}, Experimental and theoretical autocorrelation traces corresponding to the spectra shown in \textbf{c}.}
    \label{fig:results}
\end{figure*}

To confirm these predictions, we use an active fiber cavity setup~\cite{englebert_temporal_2021} which permits using lossy elements such as an isolator or a drop port while maintaining a high finesse (in the unsaturated gain regime). 
Our experimental set-up is shown in Fig.~\ref{fig:setup}a. The cavity is mostly made of standard single mode fiber [$L$=47~m, $\beta_2=-23~\text{ps}^2/\text{km}$,$\gamma=1.3\,(\text{W.km})^{-1}$]. It includes an isolator, a 90/10 through port, a 99/1 drop port, a 1535/1550 wavelength multiplexer, two 1480/1550 wavelength multiplexers, and a short erbium-doped fiber ($\sim$30~cm).
The driving laser is a highly coherent 1550~nm distributed feedback laser (linewidth $\leq$ 100~Hz), amplified to $P_d$ = 100~mW. It is sent in the cavity through the 90/10 coupler ($\theta=10\%$). The active fiber is pumped with 2~W at 1480~nm. 
We define the effective loss $\Lambda_e$ as the difference between the intrinsic roundtrip loss and the amplifier gain~\cite{englebert_temporal_2021}. 
In the experiments discussed below, the intracavity power is much lower than the saturation power ($\approx$950~mW) such that the effective loss can be considered constant $(\Lambda_e\approx2\%,\mathcal{F}\approx315)$. In that regime, the cavity behaves as a high finesse passive resonator and the scaling laws discussed above apply $(\Lambda\rightarrow \Lambda_e)$.
The main novelty in this experiment, as compared to~\cite{englebert_temporal_2021}, is the use of a frequency-shifted control signal~\cite{li_experimental_2020} which allows us to probe a wide region of detunings ($0.5<\delta<5.7$). In our previous experiment, the detuning was stabilized using the background power which limited the range to $(\delta<0.5)$.
The experiment is conducted as follows. We first stabilize the cavity at a low detuning ($\delta=0.5$), and then excite a soliton by sending a \emph{single} 1535~nm writing pulse through a wavelength multiplexer. The writing pulses are coming from a 1535~nm mode-locked laser and are gated using an acousto optic modulator~\cite{leo_temporal_2010,jang_ultraweak_2013}.
Once the soliton is excited, we change the detuning using the frequency-shifter to probe the soliton branch~\cite{englebert_temporal_2021}.
We scan up to $\delta=2.9$ in steps of 0.015 rad (at an approximate rate of 0.03 rad/s), beyond which the soliton disappears. This maximum experimental detuning ($\delta_\mathrm{exp}^\mathrm{max} = 2.9$) is much lower than the theoretical one ($\delta^\mathrm{max}=7.5$).
This is likely due to the Raman effect, which is known to limit the extension of the soliton branch ~\cite{wang_addressing_2018}.
A Raman-induced self-frequency shit of the soliton is indeed notable in Fig.~\ref{fig:results}a and b.
We will hence compare theory and experiments using the expressions \eqref{eqsol}-\eqref{eqratio}, up to the maximum experimental detuning. A detailed experimental analysis of the impact of the self-frequency shift of CSs is beyond the scope of this paper.

The 160 spectra are shown as a colormap in Fig.~\ref{fig:results}a. Two examples are plotted in Fig.~\ref{fig:results}b and their respective autocorrelation trace are shown in Fig.~\ref{fig:results}c. Both are compared to the scaling law \eqref{eqsol}.
The agreement is excellent, confirming the generation of sech-shaped pulses in the cavity. There is a small discrepancy between the analytical and experimental spectra, which is likely due to the dispersion of cavity transmission over the soliton bandwidth (see supplementary figure). 


\begin{figure}
    \centering
    \includegraphics[scale=0.9]{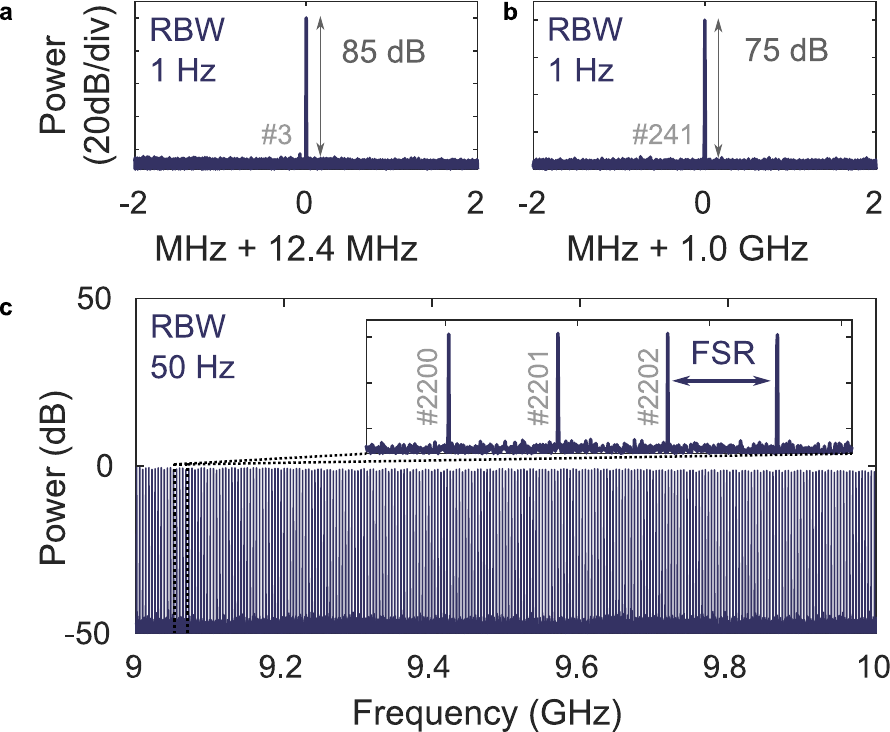}
    \caption{\textbf{a-c}, Portions of the experimental electrical spectrum showing beatnotes every free spectral range (FSR).}
    \label{fig:RF}
\end{figure}

The optical spectra do not display a comb-like structure because the repetition rate (4~MHz) is below the resolution of the optical spectrum analyzer. To confirm the presence of a single soliton in the cavity, which is important to extract the peak-to-background-ratio, we show the electrical spectrum corresponding to the pulse train obtained with $\delta=2.9$ (see Fig.~\ref{fig:RF}). We use an AC coupled 12.5~GHz photodiode connected to an electrical spectrum analyzer.
We display a 1~GHz window close to 10~GHz as well as two lower frequency sidebands. The flat spectral envelope and narrow band beat notes confirm the generation of a stable 4~MHz electrical pulse train, hence the presence of a single soliton in the cavity. Note that such a comb is potentially interesting for mm-wave spectroscopy where highly dense electric combs with narrowband beatnotes are required~\cite{tammaro_high_2015}. The FWHM of the beatnotes are here resolution limited, and the bandwidth can be increased by using faster detection. Moreover, our system can be adapted to generate soliton combs with a wide range of repetition rates. There is in principle no upper limit on the period of the pulse train (cavity length) because the system operates below the lasing threshold in the absence of solitons.

\begin{figure}
    \centering
    \vspace{2mm}
    \includegraphics[scale=0.9]{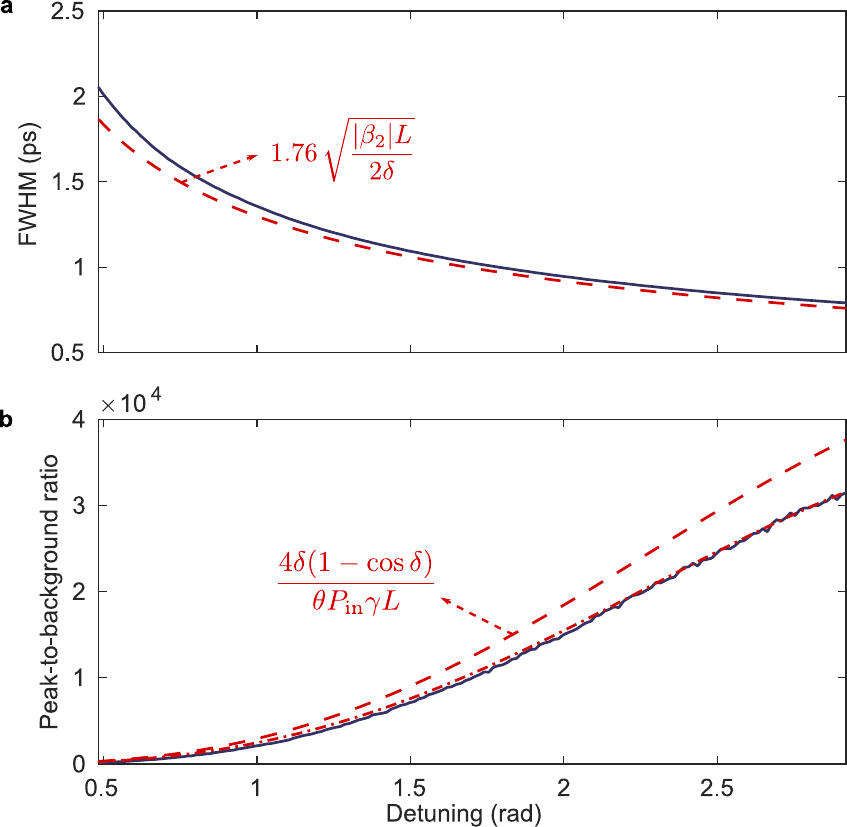}
    \caption{Experimental (solid line) and theoretical (dashed line) soliton duration (\textbf{a}) and peak-to-background ratio (\textbf{b}) as a function of the detuning. The dash-dotted line in \textbf{b} corresponds to the scaled theoretical curve (see text).}
    \label{fig:soliton}
\end{figure}

In Figure~\ref{fig:soliton}, the soliton FWHM $t_s=1.76\sqrt{\frac{|\beta_2|L}{2\delta}}$ and peak-to-background ratio are plotted as a function of the detuning and compared to experimental results.
To calculate the experimental ratio, we extract the background power $P_b$ and the soliton average power $P_s^{\mathrm{av}}$ from the spectra shown in Figure~\ref{fig:results} by integrating the spectral density over their respective bandwidth. For the soliton, the background is removed by fitting the spectrum to a $\mathrm{sech}^2$.
We then use the relation $R=\frac{P_s^{\mathrm{av}}t_r}{P_b t_s}$, where $t_r$ is the rountrip time.
We find a significant discrepancy between the experimental and theoretical ratios. The difference is as high as 40\% at $\delta=0.5$ but gradually decreases to reach a plateau of about 16\% for $\delta>2$. It may be due to intracavity dynamics, not captured by the mean field model, but in simulations of the complete Ikeda map~\cite{englebert_temporal_2021} we find less than 10\% intracavity variation of the ratio. To highlight the role played by the detuning, we scale the theoretical ratio to the experimental one in Figure~\ref{fig:soliton}. We use a constant scaling factor $K = 0.84$.
The evolution with the detuning of both the duration and the ratio matches the scaling laws and confirms the importance of the cavity detuning parameter. Going from $\delta=0.5$ to $\delta=2.9$, the soliton duration is halved and the peak-to-background ratio is increased by a factor of 150.
While the impact of the detuning on the duration of CSs has already been studied in detail (see, e.g., \cite{lucas_detuning-dependent_2017}), it is to the best of our knowledge the first report of the peak-to-background ratio dependence on the detuning. Moreover, it is the first time that stable CSs are excited close to the anti-resonance condition under continuous-wave driving. At $\delta=2.9$, the FWHM of the solitons is 780~fs. The corresponding peak power is 93~W, which is 31.000 times the measured background power ($P_b=3$~mW).
While the ratio is infinite in mode-locked lasers, we stress that CSs are phase-locked to a highly coherent continuous-wave laser (the soliton disappears if the driving is switched off). 

In conclusion, we generated a low repetition rate soliton pulse train with an unprecedented peak-to-background ratio by harnessing the cavity detuning in a coherently driven fiber laser. We find good agreement between our results and simple scaling laws describing the soliton and its background. 
The maximum detuning is limited, likely by the
Raman effect, in this experiment but higher detunings
could be reached by using longer cavities~\cite{anderson_coexistence_2017}.
Several other improvements can be foreseen. High output energies can be obtained by increasing the transmission ratio of the drop port and shorter solitons could be obtained through dispersion management~\cite{jang_observation_2014}.
Finally, we note that the same solitons, but without the continuous wave background, can in principle be generated by driving the cavity through phase-sensitive amplification~\cite{englebert_parametrically_2021}.

\section*{Acknowledgments}
This work was supported by funding from the European Research Council (ERC) under the European Union’s Horizon 2020 research and innovation program (grant agreement No 757800). N.E., C.M.A. and F.L. acknowledge the support of the "Fonds de la Recherche Scientifique" (FNRS, Belgium).

\section*{Data Availability Statement}
The data that support the findings of this study are available from the corresponding author upon reasonable request.


\bibliography{ACS_Raman} 

\begin{thebibliography}{10}
\expandafter\ifx\csname url\endcsname\relax
  \def\url#1{\texttt{#1}}\fi
\expandafter\ifx\csname urlprefix\endcsname\relax\def\urlprefix{URL }\fi
\providecommand{\bibinfo}[2]{#2}
\providecommand{\eprint}[2][]{\url{#2}}

\bibitem{leo_temporal_2010}
\bibinfo{author}{Leo, F.} \emph{et~al.}
\newblock \bibinfo{title}{Temporal cavity solitons in one-dimensional {Kerr}
  media as bits in an all-optical buffer}.
\newblock \emph{\bibinfo{journal}{Nature Photonics}}
  \textbf{\bibinfo{volume}{4}}, \bibinfo{pages}{471--476}
  (\bibinfo{year}{2010}).
\newblock \urlprefix\url{https://www.nature.com/articles/nphoton.2010.120}.

\bibitem{jang_ultraweak_2013}
\bibinfo{author}{Jang, J.~K.}, \bibinfo{author}{Erkintalo, M.},
  \bibinfo{author}{Murdoch, S.~G.} \& \bibinfo{author}{Coen, S.}
\newblock \bibinfo{title}{Ultraweak long-range interactions of solitons
  observed over astronomical distances}.
\newblock \emph{\bibinfo{journal}{Nature Photonics}}
  \textbf{\bibinfo{volume}{7}}, \bibinfo{pages}{657--663}
  (\bibinfo{year}{2013}).
\newblock \urlprefix\url{https://www.nature.com/articles/nphoton.2013.157/}.

\bibitem{englebert_temporal_2021}
\bibinfo{author}{Englebert, N.}, \bibinfo{author}{Mas~Arabí, C.},
  \bibinfo{author}{Parra-Rivas, P.}, \bibinfo{author}{Gorza, S.-P.} \&
  \bibinfo{author}{Leo, F.}
\newblock \bibinfo{title}{Temporal solitons in a coherently driven active
  resonator}.
\newblock \emph{\bibinfo{journal}{Nature Photonics}}
  \textbf{\bibinfo{volume}{15}}, \bibinfo{pages}{536--541}
  (\bibinfo{year}{2021}).
\newblock \urlprefix\url{https://www.nature.com/articles/s41566-021-00807-w}.

\bibitem{herr_temporal_2014}
\bibinfo{author}{Herr, T.} \emph{et~al.}
\newblock \bibinfo{title}{Temporal solitons in optical microresonators}.
\newblock \emph{\bibinfo{journal}{Nature Photonics}}
  \textbf{\bibinfo{volume}{8}}, \bibinfo{pages}{145--152}
  (\bibinfo{year}{2014}).
\newblock \urlprefix\url{https://www.nature.com/articles/nphoton.2013.343}.

\bibitem{yi_soliton_2015}
\bibinfo{author}{Yi, X.}, \bibinfo{author}{Yang, Q.-F.}, \bibinfo{author}{Yang,
  K.~Y.}, \bibinfo{author}{Suh, M.-G.} \& \bibinfo{author}{Vahala, K.}
\newblock \bibinfo{title}{Soliton frequency comb at microwave rates in a
  high-{Q} silica microresonator}.
\newblock \emph{\bibinfo{journal}{Optica}} \textbf{\bibinfo{volume}{2}},
  \bibinfo{pages}{1078--1085} (\bibinfo{year}{2015}).
\newblock
  \urlprefix\url{https://www.osapublishing.org/optica/abstract.cfm?uri=optica-2-12-1078}.

\bibitem{lilienfein_temporal_2019}
\bibinfo{author}{Lilienfein, N.} \emph{et~al.}
\newblock \bibinfo{title}{Temporal solitons in free-space femtosecond
  enhancement cavities}.
\newblock \emph{\bibinfo{journal}{Nature Photonics}}
  \textbf{\bibinfo{volume}{13}}, \bibinfo{pages}{214--218}
  (\bibinfo{year}{2019}).
\newblock \urlprefix\url{https://www.nature.com/articles/s41566-018-0341-y}.

\bibitem{akhmediev_dissipative_2005}
\bibinfo{editor}{Akhmediev, N.} \emph{et~al.} (eds.)
  \emph{\bibinfo{title}{Dissipative {Solitons}}}, vol. \bibinfo{volume}{661} of
  \emph{\bibinfo{series}{Lecture {Notes} in {Physics}}}
  (\bibinfo{publisher}{Springer}, \bibinfo{address}{Berlin, Heidelberg},
  \bibinfo{year}{2005}).
\newblock \urlprefix\url{http://link.springer.com/10.1007/b11728}.

\bibitem{grelu_dissipative_2012}
\bibinfo{author}{Grelu, P.} \& \bibinfo{author}{Akhmediev, N.}
\newblock \bibinfo{title}{Dissipative solitons for mode-locked lasers}.
\newblock \emph{\bibinfo{journal}{Nature Photonics}}
  \textbf{\bibinfo{volume}{6}}, \bibinfo{pages}{84--92} (\bibinfo{year}{2012}).
\newblock \urlprefix\url{https://www.nature.com/articles/nphoton.2011.345}.

\bibitem{suh_microresonator_2016}
\bibinfo{author}{Suh, M.-G.}, \bibinfo{author}{Yang, Q.-F.},
  \bibinfo{author}{Yang, K.~Y.}, \bibinfo{author}{Yi, X.} \&
  \bibinfo{author}{Vahala, K.~J.}
\newblock \bibinfo{title}{Microresonator soliton dual-comb spectroscopy}.
\newblock \emph{\bibinfo{journal}{Science}} \textbf{\bibinfo{volume}{354}},
  \bibinfo{pages}{600--603} (\bibinfo{year}{2016}).
\newblock \urlprefix\url{https://science.sciencemag.org/content/354/6312/600}.

\bibitem{trocha_ultrafast_2018}
\bibinfo{author}{Trocha, P.} \emph{et~al.}
\newblock \bibinfo{title}{Ultrafast optical ranging using microresonator
  soliton frequency combs}.
\newblock \emph{\bibinfo{journal}{Science}} \textbf{\bibinfo{volume}{359}},
  \bibinfo{pages}{887--891} (\bibinfo{year}{2018}).
\newblock \urlprefix\url{https://www.science.org/doi/10.1126/science.aao3924}.

\bibitem{englebert_bloch_2023}
\bibinfo{author}{Englebert, N.} \emph{et~al.}
\newblock \bibinfo{title}{Bloch oscillations of coherently driven dissipative
  solitons in a synthetic dimension}.
\newblock \emph{\bibinfo{journal}{Nature Physics}} \bibinfo{pages}{1--8}
  (\bibinfo{year}{2023}).
\newblock \urlprefix\url{https://www.nature.com/articles/s41567-023-02005-7}.

\bibitem{xue_super-efficient_2019}
\bibinfo{author}{Xue, X.}, \bibinfo{author}{Zheng, X.} \&
  \bibinfo{author}{Zhou, B.}
\newblock \bibinfo{title}{Super-efficient temporal solitons in mutually coupled
  optical cavities}.
\newblock \emph{\bibinfo{journal}{Nature Photonics}}
  \textbf{\bibinfo{volume}{13}}, \bibinfo{pages}{616--622}
  (\bibinfo{year}{2019}).
\newblock \urlprefix\url{https://www.nature.com/articles/s41566-019-0436-0}.

\bibitem{MasArabi:22}
\bibinfo{author}{{Mas Arab\'{i}}, C.}, \bibinfo{author}{Englebert, N.},
  \bibinfo{author}{Parra-Rivas, P.}, \bibinfo{author}{Gorza, S.-P.} \&
  \bibinfo{author}{Leo, F.}
\newblock \bibinfo{title}{Mode-locking induced by coherent driving in fiber
  lasers}.
\newblock \emph{\bibinfo{journal}{Opt. Lett.}} \textbf{\bibinfo{volume}{47}},
  \bibinfo{pages}{3527--3530} (\bibinfo{year}{2022}).
\newblock
  \urlprefix\url{https://opg.optica.org/ol/abstract.cfm?URI=ol-47-14-3527}.

\bibitem{anderson_observations_2016}
\bibinfo{author}{Anderson, M.}, \bibinfo{author}{Leo, F.},
  \bibinfo{author}{Coen, S.}, \bibinfo{author}{Erkintalo, M.} \&
  \bibinfo{author}{Murdoch, S.~G.}
\newblock \bibinfo{title}{Observations of spatiotemporal instabilities of
  temporal cavity solitons}.
\newblock \emph{\bibinfo{journal}{Optica}} \textbf{\bibinfo{volume}{3}},
  \bibinfo{pages}{1071--1074} (\bibinfo{year}{2016}).
\newblock
  \urlprefix\url{https://www.osapublishing.org/optica/abstract.cfm?uri=optica-3-10-1071}.

\bibitem{obrzud_temporal_2017}
\bibinfo{author}{Obrzud, E.}, \bibinfo{author}{Lecomte, S.} \&
  \bibinfo{author}{Herr, T.}
\newblock \bibinfo{title}{Temporal solitons in microresonators driven by
  optical pulses}.
\newblock \emph{\bibinfo{journal}{Nature Photonics}}
  \textbf{\bibinfo{volume}{11}}, \bibinfo{pages}{600--607}
  (\bibinfo{year}{2017}).
\newblock \urlprefix\url{https://www.nature.com/articles/nphoton.2017.140}.

\bibitem{li_experimental_2020}
\bibinfo{author}{Li, Z.} \emph{et~al.}
\newblock \bibinfo{title}{Experimental observations of bright dissipative
  cavity solitons and their collapsed snaking in a {Kerr} resonator with normal
  dispersion driving}.
\newblock \emph{\bibinfo{journal}{Optica}} \textbf{\bibinfo{volume}{7}},
  \bibinfo{pages}{1195--1203} (\bibinfo{year}{2020}).
\newblock
  \urlprefix\url{https://opg.optica.org/optica/abstract.cfm?uri=optica-7-9-1195}.

\bibitem{wabnitz_suppression_1993}
\bibinfo{author}{Wabnitz, S.}
\newblock \bibinfo{title}{Suppression of interactions in a phase-locked soliton
  optical memory}.
\newblock \emph{\bibinfo{journal}{Optics Letters}}
  \textbf{\bibinfo{volume}{18}}, \bibinfo{pages}{601--603}
  (\bibinfo{year}{1993}).
\newblock
  \urlprefix\url{https://www.osapublishing.org/ol/abstract.cfm?uri=ol-18-8-601}.

\bibitem{coen_universal_2013}
\bibinfo{author}{Coen, S.} \& \bibinfo{author}{Erkintalo, M.}
\newblock \bibinfo{title}{Universal scaling laws of {Kerr} frequency combs}.
\newblock \emph{\bibinfo{journal}{Optics Letters}}
  \textbf{\bibinfo{volume}{38}}, \bibinfo{pages}{1790--1792}
  (\bibinfo{year}{2013}).
\newblock
  \urlprefix\url{https://www.osapublishing.org/ol/abstract.cfm?uri=ol-38-11-1790}.

\bibitem{wang_addressing_2018}
\bibinfo{author}{Wang, Y.} \emph{et~al.}
\newblock \bibinfo{title}{Addressing temporal {Kerr} cavity solitons with a
  single pulse of intensity modulation}.
\newblock \emph{\bibinfo{journal}{Optics Letters}}
  \textbf{\bibinfo{volume}{43}}, \bibinfo{pages}{3192--3195}
  (\bibinfo{year}{2018}).
\newblock
  \urlprefix\url{https://www.osapublishing.org/ol/abstract.cfm?uri=ol-43-13-3192}.

\bibitem{tammaro_high_2015}
\bibinfo{author}{Tammaro, S.} \emph{et~al.}
\newblock \bibinfo{title}{High density terahertz frequency comb produced by
  coherent synchrotron radiation}.
\newblock \emph{\bibinfo{journal}{Nature Communications}}
  \textbf{\bibinfo{volume}{6}}, \bibinfo{pages}{7733} (\bibinfo{year}{2015}).
\newblock \urlprefix\url{https://www.nature.com/articles/ncomms8733}.

\bibitem{lucas_detuning-dependent_2017}
\bibinfo{author}{Lucas, E.}, \bibinfo{author}{Guo, H.}, \bibinfo{author}{Jost,
  J.~D.}, \bibinfo{author}{Karpov, M.} \& \bibinfo{author}{Kippenberg, T.~J.}
\newblock \bibinfo{title}{Detuning-dependent properties and dispersion-induced
  instabilities of temporal dissipative {Kerr} solitons in optical
  microresonators}.
\newblock \emph{\bibinfo{journal}{Physical Review A}}
  \textbf{\bibinfo{volume}{95}}, \bibinfo{pages}{043822}
  (\bibinfo{year}{2017}).
\newblock \urlprefix\url{https://link.aps.org/doi/10.1103/PhysRevA.95.043822}.

\bibitem{anderson_coexistence_2017}
\bibinfo{author}{Anderson, M.} \emph{et~al.}
\newblock \bibinfo{title}{Coexistence of {Multiple} {Nonlinear} {States} in a
  {Tristable} {Passive} {Kerr} {Resonator}}.
\newblock \emph{\bibinfo{journal}{Physical Review X}}
  \textbf{\bibinfo{volume}{7}}, \bibinfo{pages}{031031} (\bibinfo{year}{2017}).
\newblock \urlprefix\url{https://link.aps.org/doi/10.1103/PhysRevX.7.031031}.

\bibitem{jang_observation_2014}
\bibinfo{author}{Jang, J.~K.}, \bibinfo{author}{Erkintalo, M.},
  \bibinfo{author}{Murdoch, S.~G.} \& \bibinfo{author}{Coen, S.}
\newblock \bibinfo{title}{Observation of dispersive wave emission by temporal
  cavity solitons}.
\newblock \emph{\bibinfo{journal}{Optics Letters}}
  \textbf{\bibinfo{volume}{39}}, \bibinfo{pages}{5503--5506}
  (\bibinfo{year}{2014}).
\newblock
  \urlprefix\url{https://opg.optica.org/ol/abstract.cfm?uri=ol-39-19-5503}.

\bibitem{englebert_parametrically_2021}
\bibinfo{author}{Englebert, N.} \emph{et~al.}
\newblock \bibinfo{title}{Parametrically driven {Kerr} cavity solitons}.
\newblock \emph{\bibinfo{journal}{Nature Photonics}}
  \textbf{\bibinfo{volume}{15}}, \bibinfo{pages}{857--861}
  (\bibinfo{year}{2021}).
\newblock \urlprefix\url{https://www.nature.com/articles/s41566-021-00858-z}.

\end{thebibliography}
\bibliographystyle{naturemag}

\end{document}